\documentclass[letterpaper,twoside,english,reprint]{revtex4-2}
\usepackage{lmodern}

\usepackage[T1]{fontenc}
\usepackage[latin9]{inputenc}
\usepackage{color}
\usepackage{babel}
\usepackage{amsmath}
\usepackage{amssymb}
\usepackage{graphicx}
\usepackage[unicode=true,pdfusetitle,
 bookmarks=true,bookmarksnumbered=false,bookmarksopen=false,
 breaklinks=false,pdfborder={0 0 1},backref=false,colorlinks=true]
 {hyperref}
\hypersetup{
 linkcolor=blue, urlcolor=blue, citecolor=blue}

\makeatletter

\pdfpageheight\paperheight
\pdfpagewidth\paperwidth

\usepackage{hyperref}
\usepackage[all]{hypcap}

\makeatother

\begin{document}
\title{Extinctions as a vestige of instability: the geometry of stability
and feasibility}
\author{Stav Marcus, Ari M. Turner and Guy Bunin}
\affiliation{Department of Physics, Technion - Israel Institute of Technology,
Haifa 32000, Israel}
\begin{abstract}
Species coexistence is a complex, multifaceted problem. At an equilibrium,
coexistence requires two conditions: stability under small perturbations;
and feasibility, meaning all species abundances are positive. Which
of these two conditions is more restrictive has been debated for many
years, with many works focusing on statistical arguments for systems
with many species. Within the framework of the Lotka-Volterra equations,
we examine the geometry of the region of coexistence in the space
of interaction strengths, for symmetric competitive interactions and
any finite number of species. We consider what happens when starting
at a point within the coexistence region, and changing the interaction
strengths continuously until one of the two conditions breaks. We
find that coexistence generically breaks through the loss of feasibility,
as the abundance of one species reaches zero. An exception to this
rule\textendash where stability breaks before feasibility\textendash happens
only at isolated points, or more generally on a lower dimensional
subset of the boundary.

The reason behind this is that as a stability boundary is approached,
some of the abundances generally diverge towards minus infinity, and
so go extinct at some earlier point, breaking the feasibility condition
first. These results define a new sense in which feasibility is a
more restrictive condition than stability, and show that these two
requirements are closely interrelated. We then show how our results
affect the changes in the set of coexisting species when interaction
strengths are changed: a system of coexisting species loses a species
by its abundance continuously going to zero, and this new fixed point
is unique. As parameters are further changed, multiple alternative
equilibria may be found. Finally, we discuss the extent to which our
results apply to asymmetric interactions.
\end{abstract}
\maketitle

\section{Introduction}

Understanding species coexistence is a central outstanding problem
in ecology. A popular mathematical modeling framework of ecological
communities is to use ordinary differential equations, such as Lotka-Volterra
and consumer-resource models, where stable coexistence is represented
by a stable fixed-point of the dynamics \citep{may_theoretical_2007}.
Even in these simplified modeling approaches, understanding coexistence
can be a hard problem, as they describe many-variable nonlinear dynamics.
The conditions for a coexisting equilibrium in such models can be
broken down into two separate conditions \citep{roberts_stability_1974}:
feasibility, the requirement that all species abundances are strictly
positive, and stability of the fixed point.

There have been three main approaches in the study of the relations
between stability and feasibility.

One approach looks at a given system with fixed parameters, and asks
whether feasibility and stability are satisfied. This is usually done
statistically, by considering an ensemble of randomly sampled system
parameters (e.g. interaction strengths) and finding the probability
that the conditions are satisfied. While early focus was on stability
alone \citep{may_will_1972,may_stability_1973}, the need to address
feasibility as well was soon pointed out \citep{roberts_stability_1974}.
A recurring result in both numerical \citep{rozdilsky_complexity_2001-1,fowler_increasing_2009}
and analytical \citep{stone_google_2016,stone_feasibility_2018} approaches
is that when the number of species is large, feasibility is a more
restrictive condition than stability.

In a second approach, known as community assembly, system parameters
are again randomly sampled, and the system is then allowed to evolve
in time until it reaches a stable equilibrium. During this process,
several species may become extinct. When these are removed, by construction
the remaining species occupy a feasible and stable state. For many
species, several distinct behaviors (phases) have been observed, depending
on the statistics of the interaction strengths \citep{bunin_ecological_2017}.
In one, the system reaches a unique equilibrium, constrained by feasibility
rather than stability (the spectrum of the stability matrix is gapped).
For symmetric interactions, outside this phase the system enters a
phase where stability and feasibility break almost together (spectrum
is gapless) \citep{biroli_marginally_2018}; yet this appears to be
sensitive to interactions being exactly symmetric \citep{ros_generalized_2023}.
For asymmetric interactions, beyond the unique equilibrium phase the
system enters a chaotic phase where dynamics persist indefinitely.

In a third approach, one first finds possible interaction strengths
that can produce a stable fixed point. Then, for a given such interaction
set, one finds the carrying capacities that produce a feasible fixed
point \citep{vandermeer_interspecific_1975,svirezhev_stability_1983,rohr_structural_2014-1,saavedra_nested_2016,saavedra_structural_2017,song_guideline_2018}.
The size of the feasible region in the space of possible carrying
capacities is used as a measure for the structural stability of a
system, i.e., its insensitivity to change in parameters. We return
to compare this approach with ours in the Discussion section.

\begin{figure*}[!t]
\begin{centering}
\includegraphics[width=1\textwidth]{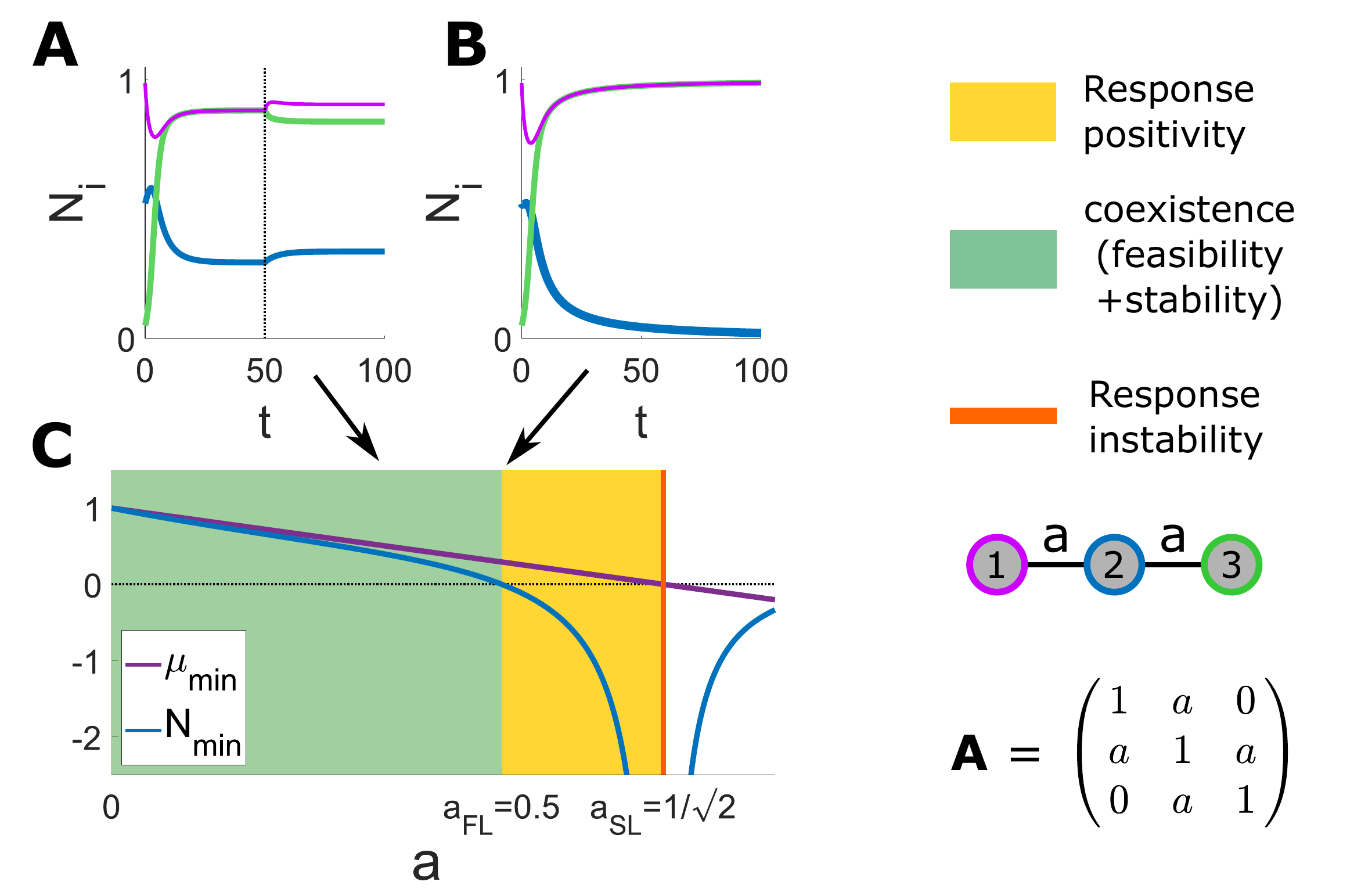}
\par\end{centering}
\caption{\label{fig:1d_example}\textbf{Feasibility, response stability, and
an example for why feasibility breaks first. }Behavior of a three
species community. The species interact with interaction strength
$a$ according to the interaction matrix $\mathbf{A}$, as shown in
the diagram above $\mathbf{A}$. \textbf{(A-B) }The species abundances
as a function of time for two values of $a$, and with arbitrarily
chosen initial conditions. Each curve represents a different species,
with the colors matching the species in the interaction diagram. (\textbf{A})
At $a=0.4$, the three-species community relaxes to an equilibrium
at time $t\le50$. The equilibrium is feasible, that is, abundance
values are positive. The equilibrium is also response-stable: At time
$t=50$ a small change to the carrying capacities is applied to the
system to test this, and produces only proportionally small changes
in the equilibrium values. Varying interaction strengths $a$, these
two conditions are satisfied for all $0<a<1/2$. (\textbf{B}) At $a=1/2$,
one species reaches zero abundance, so species no longer coexist due
to breaking of feasibility, while response stability is maintained.
(\textbf{C}) The smallest abundance $N_{\text{min}}$, and the smallest
eigenvalue of the interaction matrix $\mu_{\text{min}}$, as a function
of $a$. $\mu_{\text{min}}$ acts as a measure of response stability:
when it reaches zero at $a=a_{\text{SL}}=1/\sqrt{2}\approx0.71$,
perturbations can cause diverging changes in abundances. But we show
that this also means that the abundances themselves diverge at $a=a_{\text{SL}}$,
and in particular, $N_{\text{min}}\to-\infty$ there. This provides
a mechanism for the earlier breaking of feasibility, as $N_{\text{min}}$
must indeed cross zero already at $a<a_{\text{SL}}$.}
\end{figure*}

Here we take a different approach. We consider the space of possible
interaction strengths and consider the region where all species of
the system coexist. In this region, the conditions that must be satisfied
are feasibility and Lyapunov stability (meaning that variables return
to equilibrium if initiated close to it). We also consider another
type of stability, response stability, meaning that small changes
to carrying capacities (press perturbations \citep{bender_perturbation_1984})
only produce proportionally small changes in the equilibrium abundance
values. See an example in Fig. \ref{fig:1d_example}(A-B). Focusing
on systems with symmetric competitive interactions, we show that \emph{at
the boundaries of the coexistence region it is generically feasibility
that breaks first.} By ``generically'' we mean that this is true
everywhere except at special isolated points on the boundary (or more
generally subspaces of lower dimension). In this sense feasibility
is a more restrictive condition than stability. We proved part of
this result as a lemma in a previous paper \citep{marcus_local_2022}.
Our method is applicable to any number of species, not just ones with
asymptotically many species, in contrast to other approaches discussed
above. There might be additional interesting phenomena in the asymptotic
limit of many species besides those considered here.

The breaking of feasibility before response stability can be understood
by the following reasoning: when varying system parameters, as an
instability is approached it generically induces large variations
in species abundances. This in turn leads to extinctions before the
instability is reached, thus breaking feasibility first, see example
in Fig. \ref{fig:1d_example}(C). In this example, interactions are
changed by increasing a single parameter $a$. The range of coexistence
ends as the smallest species abundance crosses zero at a value $a=a_{\text{FL}}$,
implying loss of feasibility. This can be related to the fact that
the smallest eigenvalue of the interaction matrix reaches zero at
some larger value $a_{\text{SL}}$, which we will show is associated
with loss of stability. This causes some abundances to diverge towards
$-\infty$ at this point, and indeed cross abundance zero before it.
We will return to this example and explain it in more detail after
defining all terms in our model. This understanding highlights the
close links between stability and feasibility, and the necessity of
studying them together.

An important consequence of our results is on the set of equilibria
in ecological systems in the presence of migration, where systems
may admit multiple equilibria in which some species are extinct. We
show that as interaction strengths are changed continuously, coexistence
is lost by the abundance of a single species going to zero, and the
new fixed point where this species is extinct is unique. As parameters
are further changed, multiple equilibria may then appear.

For mutualistic symmetric interactions, we show that at the boundary
of the coexistence region feasibility and response stability break
together. In the Lotka-Volterra model that we use, abundances diverge
when crossing this boundary, indicating that this model cannot be
used in this parameter regime. For systems that have a mixture of
competitive and mutualistic interactions, either of the behaviors
above is possible, with feasibility breaking with or before response
stability. In all symmetric cases, the third possibility, where response
stability breaks before feasibility, will require a special, fine-tuned
combination of parameters. Finally, we extend our results to the case
of asymmetric interactions, and discuss when the results above still
apply.

The paper is structured as follows. We start by introducing the Lotka-Volterra
model and the two definitions of stability. We then turn to our main
claim: that for competitive symmetric interactions, feasibility generically
breaks before response stability. We first use the single parameter
example in Fig. \ref{fig:1d_example} to demonstrate the basic mechanism
behind this behavior and then prove it for a general case. We also
discuss the behavior when there are some mutualistic interactions.
Next, we discuss implications of our results on the manner in which
new fixed points are formed as interaction strengths change. Lastly,
we discuss asymmetric interactions, and when our results also extend
to this regime.

\section{The model and stability definitions}

We will consider the dynamics of $S$ species described by the Lotka-Volterra
(LV) equations, where the dynamics of the abundance $N_{i}$ of species
$i$ is given by

\begin{equation}
\frac{dN_{i}}{dt}=N_{i}g_{i}\left(\vec{N}\right)=N_{i}\rho_{i}\left(K_{i}-\sum_{j}A_{ij}N_{j}\right)\label{eq:LV}
\end{equation}
Here $g_{i}\left(\vec{N}\right)$ is the growth rate of species $i$,
that depends on the interaction strengths $A_{ij}$, the carrying
capacities $K_{i}$ (the long-time abundance reached by the species
when isolated) and the bare growth rates of the species, $r_{i}\equiv\rho_{i}K_{i}$
(the growth rate at small numbers when isolated). Here the intraspecific
interactions are all $A_{ii}=1$. For the first part of the paper,
we will take symmetric interactions, with $A_{ij}=A_{ji}$ for all
$i,j$. Both carrying capacities and bare growth rates are taken to
be positive, $K_{i},r_{i}>0$. The feasibility condition in this formulation
is that for all $i$, $N_{i}^{*}>0$, and so the fixed-point abundances
can be found immediately by setting $g_{i}\left(\vec{N}^{*}\right)=0$,
which gives $\vec{N}^{*}=\mathbf{A}^{-1}\vec{K}$, which yields a
unique fixed point that must also satisfy $N_{i}^{*}>0$ in order
for feasibility to hold. In general, all possible fixed points are
found by choosing a set of species with $N_{i}^{*}=0$, then finding
the rest of the abundances using $\vec{N}^{*}=\mathbf{A}^{-1}\vec{K}$
with $\mathbf{A},\vec{K}$ reduced to the species with $N_{i}^{*}\neq0$;
therefore once the set is chosen the fixed point is unique. Here we
will (except in section \ref{subsec:How-fixed-points}) consider the
fixed point under the assumption that all species coexist, even in
regions where it becomes unrealistic, as the solution yields some
$N_{i}^{*}<0$.

When a feasible solution exists, the system will be in equilibrium
if it is also stable. The term ``stability'' has many meanings in
ecology \citep{grimm_application_1992}. Consider first Lyapunov stability,
which is stability of the dynamics under small deviations of \emph{the
abundances} $N_{i}$ around their fixed point values $N_{i}^{*}$.
By linearizing Eq. \ref{eq:LV} about $\vec{N}^{*}$, it is satisfied
when the real parts of the eigenvalues of the matrix $\mathbf{D}(\vec{N}^{*})\mathbf{A}$
are positive, where $\mathbf{D}(\vec{N}^{*})$ is a diagonal matrix
with the equilibrium abundances $\vec{N^{*}}$ along the diagonal.
This matrix is called the community matrix \citep{may_stability_1973},
and is denoted in the following by $\mathbf{DA}$. Although Lyapunov
stability is only relevant on feasible systems (as the dynamics cannot
reach a fixed point that is not feasible), we can still formally refer
to the properties of $\mathbf{DA}$ as defining Lyapunov stability
in the entire range of interaction strengths. For symmetric interactions,
within the feasibility region the condition for Lyapunov stability
is equivalent to the interaction matrix $\mathbf{A}$ being positive
definite, denoted by $\mathbf{A}>0$ \citep{cross_three_1978,takeuchi_global_1978,berman_matrix_1983}.
Therefore it will be sufficient to consider the condition that $\mathbf{A}$
is positive definite (instead of referring to the properties of $\mathbf{DA}$)
, which is more easily accessible as it does not require finding the
abundances $N_{i}^{*}$.

This condition is related to a second type of stability, which we
term response stability. Response stability requires that once a Lyapunov
stable fixed point with abundances $\vec{N^{*}}$ is reached, a small
perturbation in the carrying capacities $K_{i}$ (known as a press
perturbation \citep{bender_perturbation_1984}) yields small changes
in the abundances $\vec{N}^{*}$ of a newly found fixed point. Since
$\partial N_{i}^{*}/\partial K_{j}=\left(\mathbf{A}^{-1}\right)_{ij}$,
response stability requires that $\mathbf{A}$ be invertible \citep{novak_characterizing_2016}.
Response instability, where $\mathbf{A}$ has a zero eigenvalue, will
first occur at the edge of the region where $\mathbf{A}>0$, which
we will call the ``response positivity'' region. Although the underlying
experiment, where a perturbation is applied, can only be done if the
system is feasible and Lyapunov stable, we will (similarly to the
case of Lyapunov stability) formally use the terms response stable
or positive to describe the properties of $\mathbf{A}$ for any values
of $A_{ij}$, even where these conditions may not be satisfied..

In the following we will consider the possible behaviors in the space
of the off-diagonal elements $A_{ij}$ of the interaction matrix,
for given $K_{i}$ and $\rho_{i}$. As $\mathbf{A}$ is an $S\times S$
matrix and $A_{ii}=1$, this space is of dimension $d=S\left(S-1\right)$
if the matrix is asymmetric, and $d=S\left(S-1\right)/2$ if it is
symmetric. It might be convenient to look at lower-dimensional cross
sections which are linear hyperplanes in this space (for example,
by taking constraints where some of the interactions are zero or some
are equal to each other, etc.). We will discuss the behavior of feasibility
and stability within these spaces of matrix components. We will first
discuss the cases where the interaction matrices $\mathbf{A}$ are
symmetric with all positive entries, and later discuss the effects
that breaking these assumptions has on the results.

\section{Feasibility breaks before stability: the basic mechanism}

To explain the basic mechanism in play, we consider an example, see
Fig. \ref{fig:1d_example}. This is a competitive (all $A_{ij}\ge0$)
three-species model, with symmetric interactions, such that the pairs
of species (1,2) and (2,3) interact with strength $a\geq0$, while
species 1 and 3 do not interact directly. The carrying capacities
and growth rates in isolation are all taken to be one, $K_{1,2,3}=1,\rho_{1,2,3}=1$.
In this simple system, one can find the minimal eigenvalue of $\mathbf{A}$,
$\mu_{\min}=1-\sqrt{2}a$, and the abundances, $N_{1}^{*}=N_{3}^{*}=\left(1-a\right)/\left(1-2a^{2}\right)$,
$N_{2}^{*}=\left(1-2a\right)/\left(1-2a^{2}\right)$. The system is
feasible for $a<1/2$ and $a>1$. It is response positive for $a<1/\sqrt{2}$,
and response stability breaks at $a=1/\sqrt{2}$ (see Fig. \ref{fig:1d_example}).
Coexistence, which requires feasibility and that $\mathbf{A}$ be
response positive, breaks along with feasibility at $a=1/2$.

Let us analyze this behavior. At $a=0$, the interaction matrix is
just $\mathbf{A}=\mathbf{I}$, with $\mathbf{I}$ the identity matrix.
Thus all $N_{i}^{*}=K_{i}=1$, since $\vec{N}^{*}=\mathbf{A}^{-1}\vec{K}=\vec{K}$,
and the system is both feasible and stable. As $a$ is increased,
the eigenvalue $\mu_{\min}$ decreases, hitting zero at $a_{\text{SL}}\equiv1/\sqrt{2}$.
How does this affect the abundances $N_{1,2,3}^{*}$? As $\mu_{\min}$
approaches zero, the entries of the inverse matrix $\mathbf{A}^{-1}$
grow in absolute value until they diverge at $a_{\text{SL}}$. Since
$\vec{N}^{*}=\mathbf{A}^{-1}\vec{K}$, the values of $\vec{N}^{*}$
will also become very large in absolute value, either positive or
negative (This is true for all but very special $\vec{K}$ vectors,
see next section.). Indeed, the minimal abundance, which is $N_{2}^{*}$
for $a<a_{\text{SL}}$, diverges to $-\infty$ precisely at $a_{\text{SL}}$.
This means that\emph{ $N_{2}^{*}$ had to cross zero at some smaller
interaction strength, $a<a_{\text{SL}}$, breaking feasibility first.}

For $a<a_{\mathrm{SL}}$, the minimum eigenvalue of the \emph{community
matrix} $\mathbf{DA}$ is $\nu_{\min}=\left(1+2a^{2}-3a\right)/\left(1-2a^{2}\right)$.
Thus, Lyapunov stability is lost along with feasibility at $a=1/2$.
As we will show below, Lyapunov stability will indeed generically
break along with feasibility. Loss of Lyapunov stability is characterized
by a slowdown of the dynamics, which occurs here when at $a=1/2$
the abundance $N_{2}^{*}$ approaches zero as a power-law in time.

To explore the special cases where stability breaks before feasibility,
we expand this simple example by allowing for the interaction between
species $(1,2)$ to have a different strength than that between species
$(2,3)$, which we denote as $a_{A}$ and $a_{B}$ respectively (see
Fig. \ref{fig:Examples of geometry}). As shown in Fig. \ref{fig:Examples of geometry},
when starting in the coexisting region and changing $a_{A},a_{B}$
continuously, response stability breaks before feasibility only along
trajectories crossing one of the points $(a_{A},a_{B})=(1,0),(0,1)$.
The behavior in this case will be expanded upon below.

\section{\label{subsec:Symmetric-competition}Feasibility generically breaks
before response stability for symmetric competition}

We now prove the main result in the general case: for symmetric matrices
with non-negative entries (i.e., competitive interactions), coexistence
is lost through loss of feasibility rather than response stability.
We start with a system in the coexistence region, by for example taking
$A_{ij}=0$ for all $i\neq j$, which has a stable fixed point with
all $N_{i}=K_{i}$. We then change the interactions along some path
in $A_{ij}$ space. We show that the feasibility condition is the
first to break, unless the choice of the path is fine tuned (namely,
goes through special points, see below).

\begin{figure}[th]
\begin{centering}
\includegraphics[width=1\columnwidth]{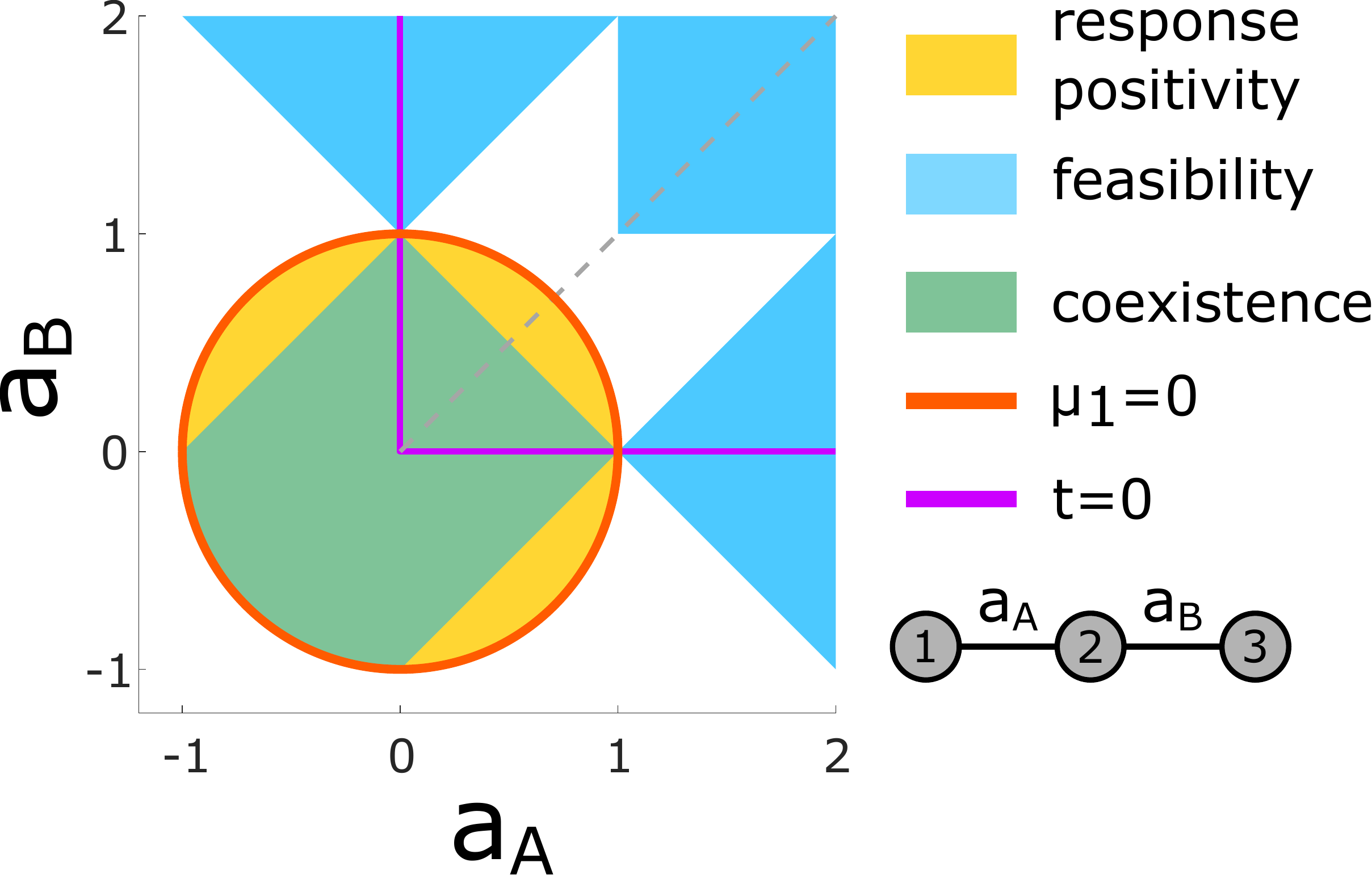}
\par\end{centering}
\caption{\label{fig:Examples of geometry}\textbf{Geometry of stability and
feasibility for symmetric interactions. }The space of interaction
strengths for a three-species system with two parameters, with the
structure of interactions as shown. For all $i$, $K_{i}=1,\rho_{i}=1$.
The dashed gray line indicates the parameter region shown in Fig.
\ref{fig:1d_example}. Feasibility, response positivity ($\mathbf{A}>0$),
and coexistence regions are shaded in blue, yellow and green respectively.
Coexistence occurs in the intersection of the response positivity
and feasibility regions. Response stability is lost on the manifold
$\mu_{1}(\mathbf{A})=0$ (orange line). As can be seen, coexistence
is lost by loss of feasibility (green region is surrounded by yellow
region), except at two points; These points lie on the manifold $t(\mathbf{A})=0$
where there are no divergences in the abundances, allowing feasibility
to be retained after loss of coexistence at the intersection of the
two manifolds.}
\end{figure}

We will start by noting that for symmetric matrices, the region where
the system is response positive with $\mathbf{A}>0$ is continuous
and convex \citep{boyd_convex_2023}, and at its edge response stability
is broken as $\mathbf{A}$ has a zero eigenvalue. However the feasibility
region may be disconnected. Indeed for the example in Fig. \ref{fig:Examples of geometry}(A-B),
the response positive region is the convex unit disk; and the feasibility
region in Fig. \ref{fig:1d_example} is disconnected (recall that
the green coexistence region is also a feasibility region).

Consider a trajectory in the $d$-dimensional space of interaction
strengths, such that as we change $\{A_{ij}\}$ continuously, response
stability is lost at a point $\mathbf{A}_{\text{SL}}$. As the matrix
$\mathbf{A}$ is symmetric, it can be diagonalized with real eigenvalues
$\left\{ \mu_{i}\left(\mathbf{A}\right)\right\} _{i=1}^{S}$, and
corresponding left- and right-eigenvectors $\vec{L}_{i}\left(\mathbf{A}\right),\vec{R}_{i}\left(\mathbf{A}\right)$,
with $\vec{L}_{i}$ being row vectors and $\vec{R}_{i}$ column vectors
(left- and right-eigenvectors are the same up to a transpose for symmetric
matrices, but are separated here to allow a generalization to some
asymmetric cases). Since response stability is lost at $\mathbf{A}_{\text{SL}}$,
at least one eigenvalue becomes zero there, which WLOG we take to
be $\mu_{1}\left(\mathbf{A}_{\text{SL}}\right)=0$. Assume for now
that there is no degeneracy, so for all $i\neq1$, $\mu_{i}\left(\mathbf{A}_{\text{SL}}\right)>0$.

The inverted matrix $\mathbf{A}^{-1}$ is written using the eigen-decomposition
of $\mathbf{A}$ as 
\begin{align}
\mathbf{A}^{-1} & =\sum_{i=1}^{S}\mu_{i}^{-1}\vec{R}_{i}\vec{L}_{i}\\
 & =\mu_{1}^{-1}\vec{R}_{1}\vec{L}_{1}+\sum_{i=2}^{S}\mu_{i}^{-1}\vec{R}_{i}\vec{L}_{i}\nonumber \\
 & \equiv\mu_{1}^{-1}\vec{R}_{1}\vec{L}_{1}+\mathbf{W}_{1}\nonumber 
\end{align}
defining the matrix $\mathbf{W}_{1}\equiv\sum_{i=2}^{S}\mu_{i}^{-1}\vec{R}_{i}\vec{L}_{i}$.
The abundances can be written as
\begin{align}
\vec{N^{*}}=\mathbf{A}^{-1}\vec{K} & =\frac{\vec{L_{1}}\vec{K}}{\mu_{1}}\vec{R}_{1}+\mathbf{W}_{1}\vec{K}\label{eq:abundances near stability break}\\
 & \equiv\frac{t_{1}\left(\mathbf{A}\right)}{\mu_{1}\left(\mathbf{A}\right)}\vec{R}_{1}+\vec{w}_{1}\nonumber 
\end{align}
with $t_{1}\left(\mathbf{A}\right)=\vec{L_{1}}\left(\mathbf{A}\right)\vec{K}$
and $\vec{w}_{1}\left(\mathbf{A}\right)=\mathbf{W}_{1}\left(\mathbf{A}\right)\vec{K}$.

In the neighborhood of $\mathbf{A}_{\text{SL}}$, all non-leading
eigenvalues (with $i\geq2$) have $\mu_{i}>0$. Therefore, the components
of $\vec{w}_{1}$ are finite, and the abundances $\vec{N^{*}}$ are
dominated by the first term in Eq. \ref{eq:abundances near stability break}.
Generically, $t_{1}\left(\mathbf{A}\right)\neq0$, and so as the trajectory
approaches $\mathbf{A}_{\text{SL}}$, $\mu_{1}\rightarrow0$ and the
abundances $N_{i}^{*}$ diverge. In a competitive system, at least
one abundance $N_{i}^{*}$ must diverge towards $-\infty$: indeed,
if for a given $i$, among the other species $j\ne i$ there are none
that diverge to $N_{j}^{*}\to-\infty$, and at least one $j$ such
that $N_{j}^{*}\to+\infty$, then $N_{i}^{*}=1-\sum_{j\ne i}A_{ij}N_{j}^{*}\rightarrow-\infty$.
By continuity of the value of $N_{i}^{*}$ along the trajectory, it
must then have $N_{i}^{*}=0$ at a point in the trajectory \emph{before}
$\mathbf{A}_{\text{SL}},$breaking feasibility. So in these cases,
\emph{feasibility must break before response stability}.

Stability can break without feasibility loss only in fine-tuned cases,
where $t_{1}\left(\mathbf{A}_{\text{SL}}\right)=0$ so the abundances
$N_{i}^{*}$ do not diverge. While the boundary of the response positivity
region is a $d-1$ dimensional manifold, $\left\{ \mathbf{A}|\min_{i}\Re\left[\mu_{i}\left(\mathbf{A}\right)\right]=0\right\} $,
feasibility is retained only on trajectories that pass through a lower
$d-2$ dimensional manifold that also satisfies $t_{i}\left(\mathbf{A}\right)=0$.
More generally, if there is a degeneracy in $\mathbf{A}$ such that
$k$ eigenvalues $\mu_{1},..,\mu_{k}$ are all zero for some region
of $\left\{ \mathbf{A}|\min_{i}\Re\left[\mu_{i}\left(\mathbf{A}\right)\right]=0\right\} $,
feasibility can be retained if the $k$ conditions $t_{i}\left(\mathbf{A}_{\text{SL}}\right)=0$
for $i=1,..k,$ are satisfied. We also note that for $t_{1}\neq0$,
there may also be fine-tuned cases where only some, and not all, of
the species abundances diverge, if by some symmetry some of the elements
of $\vec{R}_{1}$ are zero.

In the example given in Fig. \ref{fig:Examples of geometry}(A-B),
the dimension of the space of interaction strengths is $d=2$. There,
the minimal eigenvalue of $\mathbf{A}$ is $\mu_{\min}=1-\sqrt{a_{A}^{2}+a_{B}^{2}}$,
so the response positivity region is the unit disk. The corresponding
eigenvector is $\vec{L}_{1}\propto\left(a_{A},-\sqrt{a_{A}^{2}+a_{B}^{2}},a_{B}\right)$,
so $t_{1}\left(\mathbf{A}\right)=0$ along the lines $a_{A}=0$ and
$a_{B}=0$. Indeed, feasibility is retained only along trajectories
that pass through the zero-dimensional points $\left(0,1\right),\left(1,0\right)$,
where the lines $t=0$ intersect with the instability boundary.

We can also consider Lyapunov stability directly, by the condition
$\mathbf{DA}>0$, rather than through demanding both feasibility and
response positivity. One can easily see that a system is Lyapunov
stable if it is feasible and response positive; and Lyapunov unstable
if it is feasible but not response positive or vice-versa. Lyapunov
stability will therefore always break with the first among response
stability and feasibility to do so, and generically at the break of
feasibility. To see this, we use the fact that for two symmetric matrices
$\mathbf{M}_{1}$ and $\mathbf{M}_{2}$, if $\mathbf{M}_{1}$ is positive
definite, the product $\mathbf{M}_{1}\mathbf{M}_{2}$ has real eigenvalues,
with the same number of positive and negative eigenvalues as $\mathbf{M}_{2}$
\citep{serre_matrices_2010}. As one of the response stability and
feasibility conditions breaks, then respectively either $\mathbf{A}$
or $\mathbf{D}$ will have negative eigenvalues while the other matrix
remains positive definite, so $\mathbf{DA}$ will have some negative
eigenvalues and the system will not be Lyapunov stable.

When interactions are completely mutualistic, i.e., $A_{ij}<0$ for
all $i\neq j$, we prove in appendix \ref{subsec:Mutualism - response stability and feasibility}
that feasibility always breaks \emph{along with} both types of stability.
This occurs as abundances remain positive along the trajectory until
reaching the instability at $\mathbf{A}_{\text{SL}}$, where all divergences
are towards $+\infty$; but approaching $\mathbf{A}_{\text{SL}}$along
the path in the opposite direction they all diverge towards $-\infty$.
In cases where there is a mixture of both positive and negative interactions,
either case is possible: without fine-tuning, feasibility will break
either before response stability, if some of the abundances diverge
towards $-\infty$ at $\mathbf{A}_{\text{SL}}$, or along with it
if all abundance divergences are towards $+\infty$. Lyapunov stability
will again break along with the first of the other two conditions
to do so, by the same mechanism as in the all-competitive or all-mutualistic
case, depending on whether feasibility breaks before or along with
response stability respectively.

\section{\label{subsec:How-fixed-points}New fixed points are formed upon
exiting the coexistence region by a single species becoming extinct}

Here we will discuss the consequences of our results for the possible
fixed points of the system, in the case where there is a slow migration
into the system from an external species pool. Here it will be interesting
to investigate the multiple alternative equilibria of one system.
As discussed below, when migration is small, species in each equilibrium
are partitioned into ``surviving'' and ``extinct'', the latter
only supported at positive abundance by the migration. For symmetric
interactions, we discuss three results: (1) In the coexistence region,
the fully-coexisting fixed point is the only stable equilibrium of
the system; (2) For competitive systems, just after coexistence is
broken, there is a single fixed point with extinct species if this
occurs due to feasibility loss, and multiple fixed points if it is
due to stability loss; and (3) For mutualistic interactions, rather
than reaching a fixed point, abundances grow to infinity with the
dynamics outside the coexistence region and the Lotka-Volterra description
breaks down. We will here discuss points (1) and (2), and prove point
(3) in appendix \ref{subsec:Mutualism - response stability and feasibility}.

\begin{figure}[th]
\begin{centering}
\includegraphics[width=1\columnwidth]{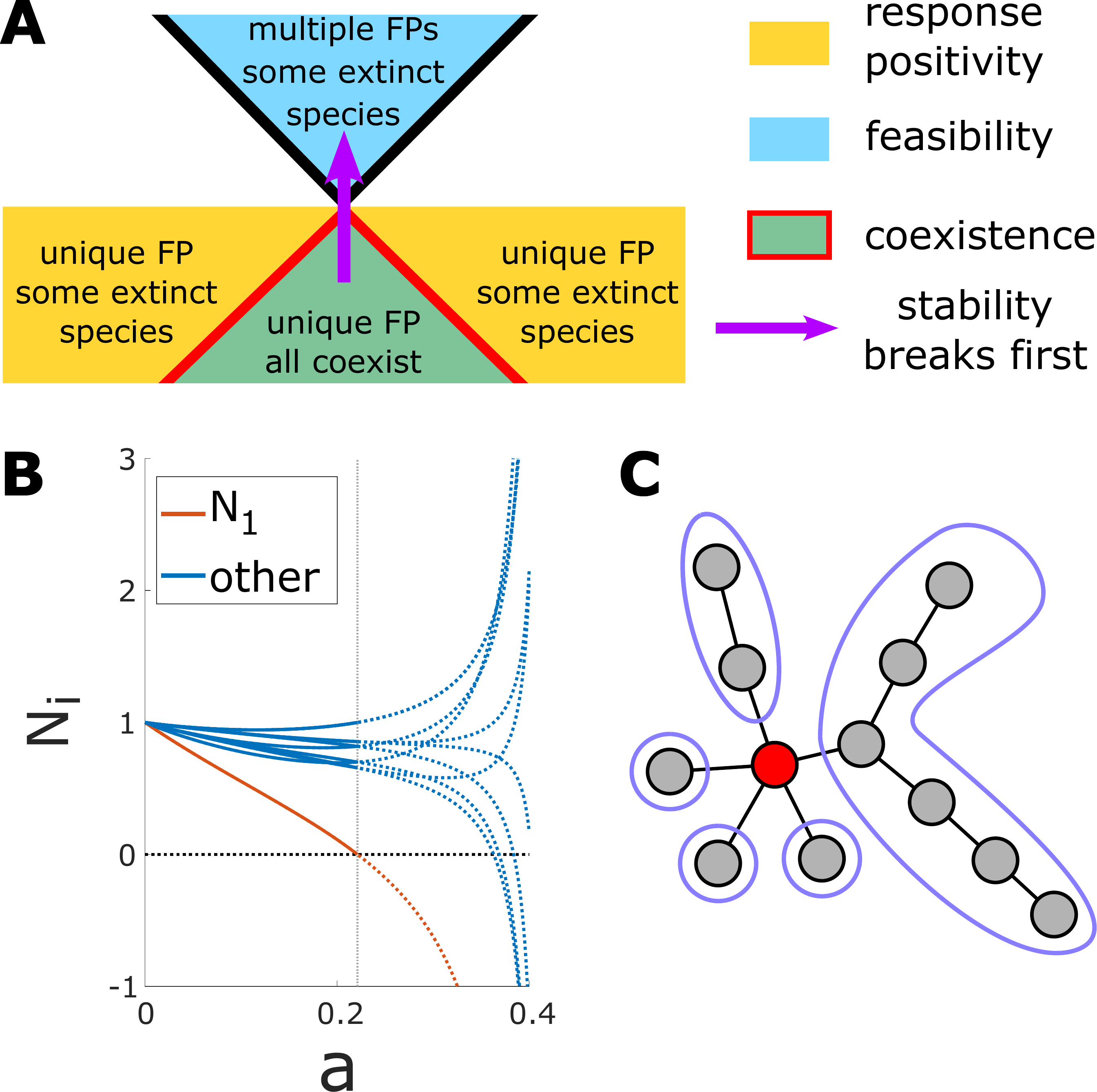}
\par\end{centering}
\caption{\label{fig:Fixed points}\textbf{Fixed points in the presence of migration.
(A)} The behavior of fixed-points at the edge of the coexistence region
in competitive systems. Feasibility, response positivity ($\mathbf{A}>0$)
and coexistence regions are shaded in blue, yellow and green respectively.
In the coexistence region, the only fixed point is the coexisting
one. When coexistence breaks and the system remains response positive,
there is a unique fixed point with extinct species. In the fine tuned
cases where stability breaks before feasibility, there will be multiple
fixed points with extinct species. \textbf{(B-C)} Example for a symmetric
system where all interaction strengths are either $A_{ij}=0$ or $A_{ij}=a$.
\textbf{(B)} The abundances as calculated by $\vec{N}^{*}=\mathbf{A}^{-1}\vec{K}$
as a function of $a$. The system coexists for $a=0$, and as $a$
is increased it loses feasibility at $a_{\text{FL}}\approx0.22$ by
the abundance of a single species, $N_{1}$, becoming zero. The calculation
assumes all species coexist, so it becomes unrealistic for $a>a_{\text{FL}}$.
\textbf{(C) }The structure of the equilibrium beyond $a_{\text{FL}}$.
The species $N_{1}$ that goes extinct at the loss of feasibility
is marked in red. In this case, when coexistence is lost, the interaction
graphs breaks into several disconnected subgraphs, shown surrounded
by purple lines.}
\end{figure}

Migration models a situation where individuals of different species
can arrive from other locations in space that are not explicitly modeled.
The Lotka-Volterra equations with constant migration read
\begin{equation}
\frac{dN_{i}}{dt}\equiv N_{i}g_{i}\left(\vec{N}\right)+\lambda_{i}\label{eq:LV with migration}
\end{equation}
The effect of adding a migration term is that it allows an extinct
species that has $g_{i}\left(\vec{N}\right)>0$ to return to the system.
At the same time, we assume here that migration rates $\lambda_{i}$
are much smaller than the other parameters in the system, formally
$\lambda_{i}\to0^{+}$. Here we will consider fixed points where some
species may be extinct with $N_{i}=0$. At a stable equilibrium, feasibility
and Lyapunov stability must hold within the set of species that are
not extinct. The calculation of the abundances $N_{i}^{*}$ and of
the stability conditions within this set can be done by taking $\lambda_{i}\approx0$.
For extinct species migration entails an added condition, known as
uninvadability, requiring that they all have negative growth rates.

For symmetric matrices, if long term abundances are bounded Eq. \ref{eq:LV with migration}
always reaches a stable fixed point, albeit possibly one where some
of the species are extinct ($N_{i}=0$). This is because the system
admits a Lyapunov function whose maxima are the fixed points of the
system \citep{macarthur_species_1970}. For all positive interactions,
abundances are bounded so the dynamics reach a fixed point with finite
abundances. If some interactions are negative, abundances can grow
until they diverge without reaching a fixed point.

Importantly, the Lyapunov function is convex if and only if the system
is response positive \citep{pykh_lyapunov_2001}. Therefore, if the
system is response positive the Lyapunov function cannot have multiple
minima, so there is a unique stable fixed point. From here one immediately
sees that in the coexistence region, where the system is response
positive, the fixed point where all species coexist is the \emph{only}
equilibrium of the system \citep{marcus_local_2022}. From our main
result, as parameters are changed, on approaching the edge of the
coexistence region, the abundance of a single species generically
goes continuously to zero while the system remains response positive.
The system will then have a unique fixed point with only this species
extinct, with abundances reached continuously from those at the coexisting
fixed point.

An example application of this result, discussed in \citep{marcus_local_2022},
is of systems with many species and sparse interactions, meaning that
most $A_{ij}=0$. Such systems can be represented as sparse graphs,
with vertices representing species and edges connecting interacting
species. For large interaction strengths, at a fixed point the system
breaks into many subsystems of persistent species (with $N_{i}>0$),
separated by extinct species, and which can be considered as isolated
systems. As interaction strengths are increased further, when any
of these susbsystems stops coexisting it will be by a single species
becoming extinct. Graphically, the subgraph of persistent species
will further break into smaller graphs by removal of one extinct species,
see example in Fig. \ref{fig:Fixed points}B.

Without fine-tuning, a coexisting system cannot reach a phase with
multiple fixed points by an infinitesimal change of parameters. This
is because the occurrence of multiple fixed points requires stability
to break down, which will occur only when varying the parameters further
from the coexistence region, so that stability breaks as well. In
the fine-tuned cases where stability breaks before feasibility, the
system can have multiple fixed points, and in fact always has them
just outside the coexistence region, as we prove in \ref{subsec:FP appendix}.
Further, the abundances at these new fixed points will not be reached
continuously from those at the coexisting fixed point, as by definition
they must be far from the unstable feasible fixed point. The geometry
of the region near such a special point is shown in Fig. \ref{fig:Fixed points}A,
showing the fully-coexisting region, regions with partial coexistence
(some species extinct), and multiple equilibria.

\section{\label{subsec:asymmetric-matrices}Asymmetric interactions}

In this section we will discuss systems with asymmetric competitive
interactions, i.e., where $A_{ij}\neq A_{ji}$ for some $i\ne j$,
and all $A_{ij}\geq0$. We start by discussing how the definitions
we have used for response and Lyapunov stability are generalized to
the asymmetric case. As in the symmetric case, we will see that generically
feasibility breaks before response stability as a result of divergences
in abundances. However in a difference from the symmetric case, coexistence
can be constrained by Lyapunov stability alone, which would break
before either feasibility or response stability without any fine tuning.
We will discuss and show examples of cases where Lyapunov stability
indeed breaks first, with no break in feasibility. On the other hand,
we will show that there are classes of systems where the picture from
the symmetric case holds, and feasibility is the factor constraining
coexistence. We will then conclude with a short discussion of asymmetric
mutualistic interactions.

An important difference between the symmetric and asymmetric case
is that the matrices $\mathbf{A}$ and $\mathbf{DA}$ may have complex
spectra, with eigenvalues that are complex conjugate pairs. For asymmetric
interactions, Lyapunov stability requires that the \emph{real parts}
of all eigenvalues of $\mathbf{DA}$ be positive, causing abundances
around the fixed point values $N_{i}^{*}$ to relax to the fixed point.
In this case, even within the feasibility region, if the real parts
of the eigenvalues of $\mathbf{A}$ are positive, this does not imply
that the real parts of the $\mathbf{DA}$ eigenvalues are positive
(although this does hold for certain large random matrices with probability
one \citep{stone_feasibility_2018}). Therefore, unlike the symmetric
case, in order to understand Lyapunov stability it is not enough to
consider only the behavior of the matrix $\mathbf{A}$. Now consider
how response stability (which again is defined at a Lyapunov-stable
equilibrium) behaves in an asymmetric system. Response stability breaks
when $\partial N_{i}^{*}/\partial K_{j}=\left(\mathbf{A}^{-1}\right)_{ij}$
diverges, which happens when $\mathbf{A}$ has an eigenvalue that
is \emph{exactly} zero. In particular, unlike for Lyapunov stability,
response stability is not lost if a complex conjugate pair of $\mathbf{A}$
eigenvalues becomes purely imaginary as their real parts become zero.

As in the symmetric case, for all-positive interactions, the loss
of response stability by an eigenvalue reaching zero will generically
cause divergences in some abundances towards $-\infty$, so \emph{feasibility
will break before response stability}. Consider again changing the
interaction strengths along a path in $A_{ij}$ space, as in Sec.
\ref{subsec:Symmetric-competition}. Again defining $\mathbf{A}_{\text{SL}}$
as the point in the path where the first eigenvalue of $\mathbf{A}$
becomes zero, from Eq. \ref{eq:abundances near stability break},
if $t_{1}\left(\mathbf{A}\right)\neq0$ there must be divergences
in the abundances at the approach to $\mathbf{A}_{\text{SL}}$, meaning
that feasibility breaks earlier. Note that if the eigenvalue is complex,
and it is only its \emph{real part} that becomes zero, response stability
does not break and there would be no divergences in the abundances.

\begin{figure}[th]
\begin{centering}
\includegraphics[width=1\columnwidth]{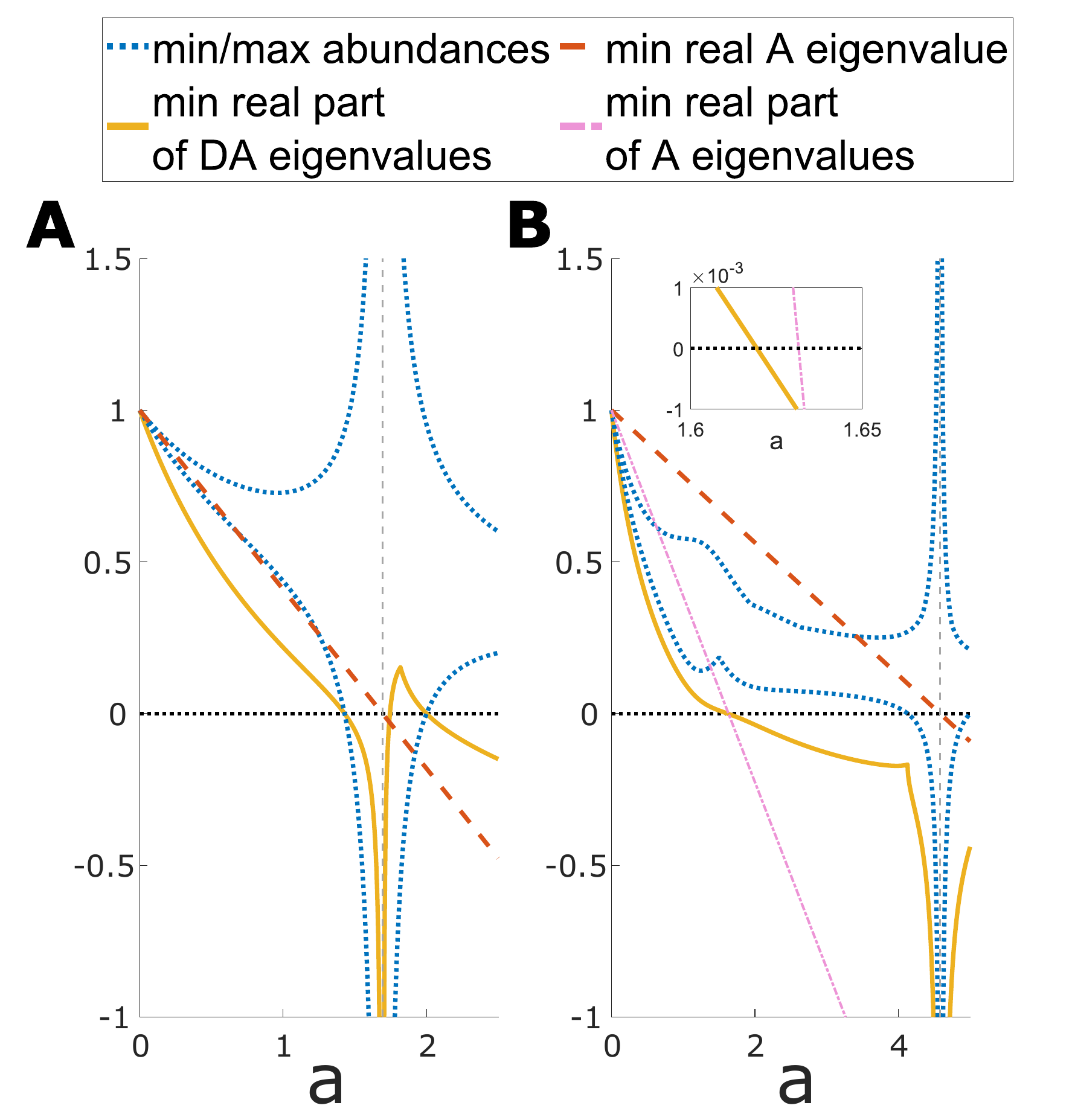}
\par\end{centering}
\caption{\textbf{\label{fig:asymmetric matrices}Asymmetric interactions. }Matrices
are of the form $\mathbf{A}=\mathbf{I}+a\mathbf{M}$, with values
shown as a function of $a$. Shown are the maximal and minimal abundances,
the minimum of the real parts of $\mathbf{DA}$ eigenvalues, and the
smallest of the real-valued eigenvalues of $\mathbf{A}$. A gray vertical
line marks the value of $a$ where the last becomes zero. As the curves
represent minimal/maximal values, cusps appear when different values
(say, the abundances of two different species) cross. The matrices
$\mathbf{M}$ are given in appendix \ref{sec:more asymmetric interactions }.
\textbf{(A)} The leading eigenvalue of $\mathbf{M}$ is real and the
behavior is the same as for symmetric matrices. Abundances diverge
as the eigenvalues of $\mathbf{A}$ approach zero, some of them towards
$-\infty$; Lyapunov stability is broken along with feasibility\textbf{
(B)} The leading eigenvalues of $\mathbf{M}$ are a complex conjugate
pair. Lyapunov stability breaks first without a break in response
stability, feasibility breaks later, and finally $\mathbf{A}$ becomes
uninvertible. The minimum of the real parts of $\mathbf{A}$ eigenvalues
is also shown in a thin dashed-dotted pink line, showing that there
are no divergences when it hits zero. Inset: The same results, zoomed
in to show that the minimum real parts of the eigenvalues of $\mathbf{A}$
and $\mathbf{DA}$ (pink and yellow lines, respectively) do not cross
zero at the same value of $a$.}
\end{figure}

It is possible, however, that \emph{Lyapunov} stability will break
down before feasibility when interactions are asymmetric. Recalling
the situation in the symmetric case, starting at the coexisting region
and changing $A_{ij}$ continuously, either feasibility or response
stability had to break down in order for Lyapunov stability to do
so. As the eigenvalues of $\mathbf{DA}$ are real in the coexisting
region, Lyapunov stability is lost by $\det(\mathbf{DA})=0$, which
means that either $\det\mathbf{A}=0$, implying the loss of response
stability, or $\det\mathbf{D}=0$, implying loss of feasibility. However
if interactions are asymmetric this is not generally true, as Lyapunov
stability can be lost by a pair of complex conjugate eigenvalues crossing
the imaginary axis, so that $\det(\mathbf{DA})\neq0$. There can therefore
be cases where Lyapunov stability is lost before either feasibility
or response stability, and so Lyapunov stability would be the condition
limiting coexistence.

One example where Lyapunov stability is the more constraining condition
without fine tuning, is cases where interaction strengths are increased
indefinitely without loss of feasibility, a scenario which is impossible
for symmetric interactions. This can occur along a trajectory in the
$\left\{ A_{ij}\right\} $ space where the smallest real eigenvalue
of $\mathbf{A}$ is always positive, so $\mathbf{A}$ is always invertible
and $\vec{N}^{*}=\mathbf{A}^{-1}\vec{K}$ never diverges. In some
of these cases, abundances never reach zero for any finite-valued
interaction strengths. On the other hand, Lyapunov stability always
breaks for large enough interaction strengths, constraining coexistence.
See Appendix \ref{sec:more asymmetric interactions } for further
discussion and examples.

On the other hand, there are several interesting cases where, as in
the symmetric case, feasibility is still the condition constraining
coexistence:
\begin{enumerate}
\item The leading eigenvalue (meaning that with the smallest real part)
of $\mathbf{DA}$ is real. This means that the smallest eigenvalue
of $\mathbf{DA}$ has to cross zero in order to lose Lyapunov stability,
and so the argument from the symmetric case applies. Our main result
for symmetric interactions is therefore robust up to a finite asymmetric
perturbation: consider a symmetric interaction matrix $\mathbf{A}_{\text{S}}$,
to which asymmetry is added by taking $\mathbf{A}=\mathbf{A}_{\text{S}}+\epsilon\mathbf{A}_{\text{AS}}$,
where $\mathbf{A}_{\text{AS}}$ is an asymmetric matrix and $\epsilon\in\mathbb{R}$.
When changing $\epsilon$ continuously from zero, there will be some
finite region where the leading eigenvalue of $\mathbf{DA}$ remains
real, as the eigenvalues would change continuously with $\epsilon$
and the appearance of a complex conjugate pair of eigenvalues would
require two eigenvalues to meet on the real axis.
\item The leading eigenvalue of $\mathbf{A}$ is real. Testing linear trajectories
in $\left\{ A_{ij}\right\} $ space that cross the origin, we find
that losing Lyapunov stability before feasibility seems to be a rare
occurrence in these cases (see appendix \ref{subsec:Lyapunov-stability-real leading EV}),
although this may change for different types of trajectories.
\item The interaction graph is tree-like (i.e., it has no loops). In this
case, the interactions can be symmetrized by rescaling the abundances
as $N_{i}\rightarrow n_{i}=N_{i}/\tilde{K}_{i}$ with some $\tilde{K}_{i}>0$
(so that the carrying capacity becomes $K_{i}/\tilde{K}_{i}$) \citep{pykh_lyapunov_2001}.
The $\left\{ N_{i}\right\} $ system is at a fixed point iff $\left\{ n_{i}\right\} $
is at a fixed point, and their stability and feasibility are the same
(as $n_{i}>0$ iff $N_{i}>0$). Therefore, the stability and feasibility
regions of the $\left\{ n_{i}\right\} $ system are linear transformations
of the regions of the $\left\{ N_{i}\right\} $ system whose geometry
remains the same.
\end{enumerate}
In Fig. \ref{fig:asymmetric matrices} we show examples of possible
behaviors for asymmetric interactions, using linear paths in $\left\{ A_{ij}\right\} $
space: we take $\mathbf{A}=\mathbf{I}+a\mathbf{M}$, with $\mathbf{I}$
the identity matrix, $\mathbf{M}$ a constant matrix with zeros on
the diagonal, and $a>0$ a real free parameter. The behaviors are
not fine-tuned, and remain unchanged when adding random perturbations
to $M_{ij}$, $i\neq j$, from a normal distribution $\mathcal{N}\left(0,10^{-4}\right)$.
For the matrices $\mathbf{A},\mathbf{DA}$ the relevant value for
the loss of response/Lyapunov stability are shown: for $\mathbf{A}$
the smallest eigenvalue that is real, and for $\mathbf{DA}$ the smallest
of the real parts of the eigenvalues. In \ref{fig:asymmetric matrices}A,
all eigenvalues of $\mathbf{A}$ are real and $\mathbf{DA}$ has a
real leading eigenvalue in the range shown, so the resulting behavior
is the same as for symmetric interactions where feasibility breaks
first. The abundances diverge as the minimal eigenvalue of $\mathbf{A}$
reaches zero, and so before this happens feasibility is lost and the
minimum real part of the $\mathbf{DA}$ eigenvalues becomes zero.
In \ref{fig:asymmetric matrices}B, neither the leading eigenvalue
of $\mathbf{A}$ nor of $\mathbf{DA}$ is real. Lyapunov stability
breaks first, followed by feasibility, and lastly an eigenvalue of
$\mathbf{A}$ becomes zero. At this point, the abundances (of the
Lyapunov unstable fixed point) diverge. Note that there is no divergence
in the abundances when the \emph{minimal real part} of the $\mathbf{A}$
eigenvalues becomes zero.

We conclude this section with a short discussion of the case of asymmetric
all-mutualistic interactions (all $A_{ij}\leq0$). If the interaction
matrix $\mathbf{A}$ is irreducible (i.e., the associated interaction
graph is strongly connected), feasibility and response stability break
together, see proof in appendix \ref{subsec:Mutualism - response stability and feasibility}.
In addition, for all-mutualistic interactions, when the system is
feasible the eigenvalues of $\mathbf{DA}$ have positive real parts
if the eigenvalues of $\mathbf{A}$ have positive real parts \citep{cross_three_1978,takeuchi_global_1978,berman_matrix_1983}.
Lyapunov stability will therefore hold at least up to the point where
feasibility breaks. The condition of irreducibility is one which we
can reasonably expect to hold in ecological systems. For example,
it holds for cases where interactions are bidirectional, i.e. where
$A_{ij}\neq0\iff A_{ji}\neq0$, as long as the system is not made
up of separated components (in which case, each component can be considered
as a separate system).

\section{Discussion}

We have considered, within the framework of the Lotka-Volterra equations,
the relations between the feasibility and stability conditions required
for species coexistence. For communities with symmetric competitive
interactions, we find that as interaction strengths are changed continuously,
coexistence is generically broken due to loss of feasibility. Stability
can break before feasibility only in fine-tuned cases, and small changes
in parameters will lead to feasibility breaking first. In this sense,
feasibility is the condition restricting coexistence, so that stability
alone is insufficient to understanding coexistence. While this was
previously discussed for many-variable systems with randomly drawn
interactions \citep{stone_feasibility_2018}, we show this here for
any finite number of species. Furthermore, feasibility and stability
are closely interlinked: the loss of feasibility happens because abundances
diverge when approaching response instability. This result is also
robust to some finite asymmetric perturbations of the interactions.

Consider for comparison the approach of \citep{svirezhev_stability_1983,logofet_matrices_1993,rohr_structural_2014-1}
to the stability-feasibility relation. In those works the two conditions
are separated by taking the carrying capacities $K_{i}$ as free parameters.
One can then first choose $a_{ij}$ such that the system is stable,
then choose $K_{i}$ values that yield a feasible fixed point, which
will give the feasibility region $\left\{ \vec{K}|\vec{K}=\mathbf{A}^{-1}\vec{N},N_{i}>0\right\} $.
This is complementary to our study, which looks at the effects of
varying the interactions $\mathbf{A}$ rather than the carrying capacities
$K_{i}$. To see the relation to the behavior we describe close to
the loss of stability, note that at the loss of stability, the interaction
matrix $\mathbf{A}$ becomes singular and the feasibility region of
$\vec{K}$ collapses from an $S$-dimensional subspace to a lower-dimensional
space, and sensitivity to changes in $\vec{K}$ is infinite (loss
of structural stability). This relates to our result: retaining feasibility
close to the loss of stability requires fine-tuning.

We also studied the implications of our results to the formation of
multiple equilibria, in systems subject to external migration. Generically,
when changing parameters away from the coexisting region, a new fixed
point will appear by the abundance of a single species going continuously
to zero. With a further change of the parameters, one may then cross
into the multiple equilibrium regime.

Our work is focused on the case of a finite number of species. The
present framework may still be useful in the limit of infinitely many
species, but should be treated with care (for example, as the spectrum
of the matrices becomes continuous). In particular, our framework
can be useful in the study of large random systems in cases that are
not covered by probabilistic arguments. Such a case is found in sparse
systems, where for strong enough interactions, the surviving species
may form multiple small interacting networks, that do not interact
with each other \citep{marcus_local_2022}.

\appendix
\twocolumngrid

\section{\label{subsec:Mutualism - response stability and feasibility}Mutualistic
systems}

Here we will prove a few claims from the main text on fully mutualistic
interactions ($A_{ij}\leq0$ for all $i\neq j$). All cases discussed
here assume that the interaction matrix $\mathbf{A}$ is irreducible,
a condition likely to hold for most ecological systems: it includes
all systems where interactions are bidirectional (i.e., $A_{ij}\neq0\iff A_{ji}\neq0$),
and specifically symmetric systems, that cannot be divided into non-interacting
communities (in which case each system can be treated separately)
\citep{gantmacher_theory_2009}. First, we show that feasibility and
response stability \emph{always} break together. We then show that
Lyapunov stability holds \emph{at least} until feasibility and response
stability break, and that for symmetric interactions it breaks simultaneously
with them. Finally, we show that for the \emph{dynamics} in the presence
of migration, beyond the coexistence region the abundances all grow
to infinity.

\subsection{Feasibility and response stability break together}

\begin{figure}[th]
\begin{centering}
\includegraphics[width=1\columnwidth]{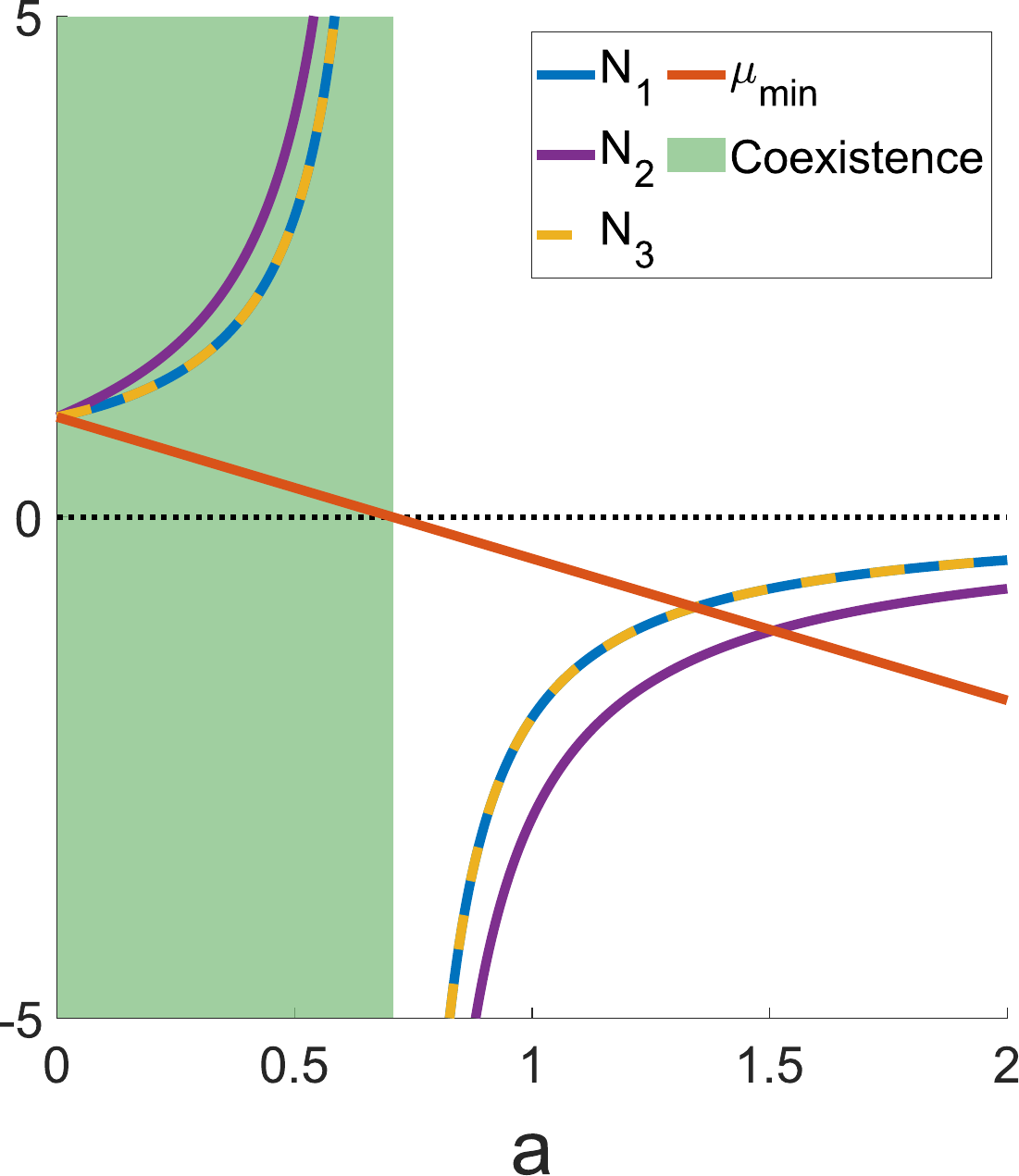}
\par\end{centering}
\caption{\textbf{\label{fig:abundances divergence mutualism}Divergences in
abundances in mutualistic systems.} Abundances and eigenvalues in
a three species system, with species (1,2) and (2,3) interacting with
$-a$, and taking $K_{i}=1,\rho_{i}=1$. Stability breaks as the minimum
eigenvalue $\mu_{\text{min}}=1-\sqrt{2}a$ (red) becomes zero at $a_{\text{SL}}=1/\sqrt{2}$.
All three abundances diverge towards $+\infty$ when approaching $a_{\text{SL}}$
from the coexistence region (green shading), and towards $-\infty$
when approaching $a_{\text{SL}}$ from outside the coexistence region.}
\end{figure}

We will now prove that for mutualistic interactions where the interaction
matrix is irreducible, feasibility and response stability break together.
The proof is structured as follows:
\begin{enumerate}
\item From \citep{su_interaction_2016}, in mutualistic systems (symmetric
or otherwise) feasibility cannot break before response stability.
To show that both break together, we need only show that there are
species with negative abundances \emph{just outside the stability
region}.
\item To do this we first show that the leading eigenvalue of $\mathbf{A}$
is real and non-degenerate, and its corresponding eigenvector has
all positive elements.
\item We use (2) to prove that on approaching $\mathbf{A}_{\text{SL}}$
from within the stable region, all abundances diverge towards $+\infty$,
and on approaching it from the unstable region, all divergences are
towards $-\infty$. Therefore all abundances are negative just outside
the stability region and the conditions break together. See example
in Fig. \ref{fig:abundances divergence mutualism}.
\end{enumerate}

To prove point (2), we use the Perron-Frobenius theorem for irreducible
non-negative matrices \citep{lemmens_nonlinear_2012}, for the matrix
$\mathbf{B}=-\left(\mathbf{A}-\mathbf{I}\right)$, where $\mathbf{I}$
is the identity matrix. $\mathbf{B}$ is non-negative: as all $A_{ii}=1$,
it has zeros along the diagonal, and as interactions are mutualistic
all other elements are non-negative. And if $\mathbf{A}$ is irreducible,
then so is $\mathbf{B}$. The theorem states that the eigenvalue of
$\mathbf{B}$ which has the maximal real part, called the Perron-Frobenius
eigenvalue $\mu_{PF}$, is real, positive and non-degenerate. Further,
its corresponding eigenvector has all positive components. The \emph{smallest}
eigenvalue of the matrix $\mathbf{A}$ is therefore $\mu_{1}=1-\mu_{PF}$,
with the same corresponding eigenvector.

To prove (3), consider again the equation for the abundances from
the main text, 
\begin{equation}
\vec{N}=\left(t_{1}/\mu_{1}\right)\vec{R}_{1}+\vec{w}_{1}\,.\label{eq: abundances near instability}
\end{equation}
From (2), $\vec{L}_{1},\vec{R}_{1}$ have all positive components,
and so also $t_{1}\left(\mathbf{A}\right)=\sum_{i}L_{1}^{i}K_{i}>0$
(recalling that all $K_{i}>0$). As instability is approached and
$\mu_{1}\rightarrow0^{+}$, the first term in Eq. \ref{eq: abundances near instability}
is strictly positive and diverges at the limit, with all other terms
finite. Therefore as stability is lost, all abundances diverge to
$N_{i}\rightarrow+\infty$. Approaching the same point from outside
the stability region, $\mu_{1}\rightarrow0^{-}$, and the abundances
are still dominated by the term $\left(t_{1}/\mu_{1}\right)\vec{R}_{1}$.
As the eigenvector components are still all positive and $\mu_{1}<0$,
all abundances diverge to $-\infty$, so the system is unfeasible.
Note that $t_{1}\left(\mathbf{A}\right)\neq0$, so there can be no
fine-tuned case where response stability breaks but feasibility does
not.

\subsection{Feasibility and Lyapunov stability break together for symmetric interactions}

We will now move to investigate Lyapunov stability. First, as mentioned
in the main text, in mutualistic systems Lyapunov stability and response
stability are equivalent in the feasible region. Lyapunov stability
therefore cannot break without a break in either feasibility or response
stability. Specifically for symmetric systems, Lyapunov stability
will break exactly at the loss of feasibility and response stability.
To see this, consider a point in the trajectory just beyond the loss
of feasibility. Here all abundances are very negative, so the matrix
$-\mathbf{D}$ is positive definite. Recall the theorem used in the
main text: for two symmetric matrices $\mathbf{M}_{1}$,$\mathbf{M}_{2}$,
if $\mathbf{M}_{1}$ is positive definite then the product $\mathbf{M}_{1}\mathbf{M}_{2}$
has the same number of positive and negative eigenvalues as $\mathbf{M}_{2}$.
Therefore, the matrix $-\mathbf{DA}$ has the same number of positive
eigenvalues as $\mathbf{A}$. As just outside the stability region
at least one eigenvalue of $\mathbf{A}$ must be positive (in the
generic case, this would be all eigenvalues except one), $-\mathbf{DA}$
has a positive eigenvalue, and therefore $\mathbf{DA}$ has at least
one negative eigenvalue and the system is not Lyapunov stable.

\subsection{Dynamics in the presence of migration for symmetric interactions}

Here we will show that outside the coexistence region abundances grow
to infinity for any initial condition, and the Lotka-Volterra description
breaks down. This is because there can be no stable and uninvadable
fixed point with extinct species, since an extinct species $i$ would
have a positive growth rate: 
\begin{align*}
g_{i}\left(\vec{N}\right) & =K_{i}-N_{i}-\sum_{j\neq i}A_{ij}N_{j}\\
 & =K_{i}+\sum_{j\neq i}\left|A_{ij}\right|N_{j}>0\,.
\end{align*}
Therefore, outside the coexistence region the system has no feasible
equilibrium nor one with extinct species. As for symmetric interactions
the system has a Lyapunov function, it cannot reach a limit cycle
or a chaotic state, and so the only possible dynamics is that some
abundances keep growing to infinity.

\section{\label{subsec:FP appendix}Scenario for generating multiple fixed
points}

\begin{figure}[th]
\begin{centering}
\includegraphics[width=1\columnwidth]{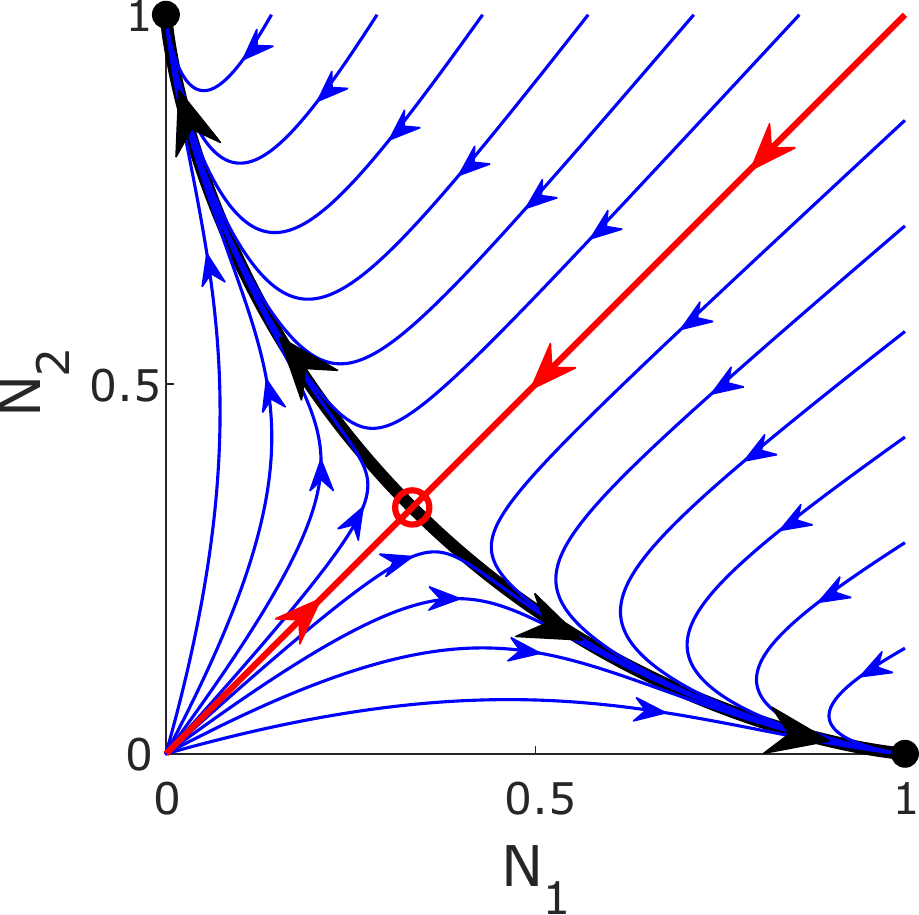}
\par\end{centering}
\caption{\textbf{\label{fig:separatrix}Two species example of the formation
of multiple fixed points in regions that are feasible and Lyapunov
unstable.} Trajectories of the LV equations in the phase space of
the abundances $\left(N_{1},N_{2}\right)$ for $A_{12}=A_{21}=2$,
and $\rho_{1,2}=1,K_{1,2}=1$. For these parameters the system is
feasible and Lyapunov unstable. The two stable fixed points $\left(N_{1},N_{2}\right)=\left(0,1\right),\left(1,0\right)$,
where one of the species is extinct, are marked with filled black
circles. The unstable coexisting fixed point $\left(N_{1}^{*},N_{2}^{*}\right)=\left(1/3,1/3\right)$,
is marked with an empty red circle. It has one stable direction, with
the trajectories of the Lotka-Volterra dynamics that stably reach
it marked in red. The unstable trajectories, starting at the unstable
fixed point and reaching the stable ones, are marked in black. Other
trajectories are shown in blue. No trajectory can cross the red line,
and therefore there must be at least one stable fixed point on either
side of it.}
\end{figure}

In this appendix we will prove the claim from the main text: for symmetric
competitive interactions, in the fine tuned cases where Lyapunov stability
is broken before feasibility, there will be multiple stable fixed
points just outside the coexistence region. To see this, we will consider
trajectories of the LV dynamics in the $S$-dimensional $\vec{N}$
space of the abundances. \emph{Note that here we are working in the
space of the abundances $N_{i}$ for some given $\mathbf{A}$, rather
than in $A_{ij}$ space considered in the rest of the paper.} For
each species, the dynamics are bounded from below by $N_{i}\geq0$.
At longer times they are also bounded from above by $K_{i}$, as the
growth rate, $K_{i}-N_{i}-\sum_{j\neq i}A_{ij}N_{j}\leq K_{i}-N_{i}$,
is negative when $N_{i}$ exceeds $K_{i}$.

As the system is still feasible, the fixed point where all species
coexist is inside this bounded region, but is unstable. As we will
show, the linear stability matrix $\mathbf{DA}$ has only a single
negative eigenvalue, so in the vicinity of the fixed point there is
\emph{a single unstable direction} of the corresponding eigenvector.
Therefore, there is an $S-1$ dimensional manifold of points that
stably reach the coexisting fixed point. See example for two species
in Fig. \ref{fig:separatrix}, where the stable manifold is a one
dimensional line in the two-dimensional $\vec{N}$ space. As trajectories
cannot cross each other, no trajectory of the LV dynamics can cross
this manifold. The stable manifold therefore separates the bounded
region into two parts, where dynamics with initial conditions in one
region cannot reach the other one. As the system has a Lyapunov function,
initial conditions in each of these two bounded regions must reach
a stable fixed point, and so in each of them there must be at least
one such fixed point. The importance of the condition that Lyapunov
stability breaks before feasibility to this argument, is that this
ensures that the coexisting fixed point is still feasible, so starting
close to it is guaranteed to lead to fixed points with positive abundances.

We need now only prove our assumption, that just outside the coexistence
region, $\mathbf{DA}$ indeed has exactly one negative eigenvalue.
To prove this, we again use the fact that the eigenvalues of a product
of two symmetric matrices, one of which is positive definite, have
the same sign as the eigenvalues of the second matrix \citep{serre_matrices_2010}.
Excluding cases with extra symmetries where $\mathbf{A}$ has some
degeneracy, response stability is lost as a single eigenvalue of $\mathbf{A}$
becomes negative. As feasibility is kept, however, all $N_{i}^{*}>0$
and the matrix $\mathbf{D}$ is positive definite. From here the linear
stability matrix $\mathbf{DA}$, like the matrix $\mathbf{A}$, has
a single negative eigenvalue.

\section{\label{sec:more asymmetric interactions }Possible behaviors for
competitive asymmetric interactions}

Here we provide additional explanations and examples of the behavior
of competitive asymmetric interactions, demonstrating that the order
in which feasibility, Lyapunov stability and response stability break
can be the same as in the symmetric case or different from it. We
expand on the scenario described in the main text, where interaction
strengths are increased without causing divergences in the abundances.
We also give the matrices used in examples in the main text. In all
examples, we consider linear paths in $\left\{ A_{ij}\right\} $ space:
we take matrices of the form $\mathbf{A}=\mathbf{I}+a\mathbf{M}$,
with $\mathbf{I}$ the identity matrix, $\mathbf{M}$ a constant non-negative
matrix with zeros on the diagonal, and $a>0$ a real, positive number,
and consider what happens as $a$ is increased.

First, the matrices used in Fig. \ref{fig:asymmetric matrices} in
the main text are 
\begin{align*}
\mathbf{M}_{\text{A}} & =\left(\begin{array}{ccc}
0 & 0.4 & 0.4\\
0.1 & 0 & 0.7\\
0.1 & 0.5 & 0
\end{array}\right)\\
\mathbf{M}_{\text{B}} & =\left(\begin{array}{cccc}
0 & 0.4 & 0.5 & 0.4\\
0.6 & 0 & 0.2 & 0.9\\
0.4 & 0.4 & 0 & 0.8\\
0.5 & 0 & 0.8 & 0
\end{array}\right)
\end{align*}
with matrix $\mathbf{M}_{\text{X}}$ the matrix used in subfigure
X. For $\mathbf{M}_{\text{A}}$ the leading eigenvalue of $\mathbf{DA}$
is real and therefore the behavior is the same as for symmetric interactions.
This is not the case for $\mathbf{M}_{\text{B}}$, where Lyapunov
stability breaks first, followed by feasibility, and lastly by response
stability.

We will now turn to the following claim from Sec. \ref{subsec:asymmetric-matrices}
of the main text: interaction strengths can in some cases be increased
indefinitely, without producing divergences in the abundances. This
occurs along paths where $\mathbf{A}$ is always invertible, so $\vec{N}^{*}=\mathbf{A}^{-1}\vec{K}$
never diverges. Feasibility may still be lost by an abundance crossing
zero (although it is not forced to do so because of a divergence),
but there are trajectories in the $A_{ij}$ space along which feasibility
is not lost for any finite interaction strength. On the other hand,
Lyapunov stability always breaks for large enough interaction strengths,
and so it will be the condition limiting coexistence in the cases
where feasibility never breaks. This behavior cannot occur for symmetric
matrices, except in the fine-tuned cases where Lyapunov stability
breaks first.

Consider again linear trajectories with $\mathbf{A}=\mathbf{I}+a\mathbf{M}$.
Consider first a symmetric matrix $\mathbf{M}$. As $\text{tr}\mathbf{M}=0$
and the eigenvalues are all real, then some must be positive and some
negative. Therefore, if $\mu_{\mathbf{M}}$ is the smallest eigenvalue
of $\mathbf{M}$, it must be negative and response stability breaks
at $a=-1/\mu_{\mathbf{M}}$. The region where a symmetric matrix is
invertible is therefore bounded, and its boundary will be crossed
along any trajectory when interactions are sufficiently increased.
Without fine-tuning, this will cause divergences in the abundances.
In contrast, if $\mathbf{M}$ is asymmetric, as it has complex eigenvalues
it might have no real negative eigenvalues even though $\text{tr}\mathbf{M}=0$.
So upon increasing $a$, response stability never breaks and there
are no divergences in the abundances.

We will now show that Lyapunov stability always breaks for large enough
$a$. Since $\mathbf{D}$ is diagonal, all $A_{ii}=1$, and the abundances
are of the order $1/a$, $\text{tr}\mathbf{DA}=\sum_{i}D_{ii}A_{ii}=\sum_{i}N_{i}=O\left(1/a\right)$.
On the other hand, as $\mathbf{A}$ is of order $a$, the off diagonal
elements of $\mathbf{DA}$ are of order $1$. Therefore, the eigenvalues
of $\mathbf{DA}$ are of order $1$, and so some of them must have
negative real parts.

\begin{figure}[th]
\begin{centering}
\includegraphics[width=1\columnwidth]{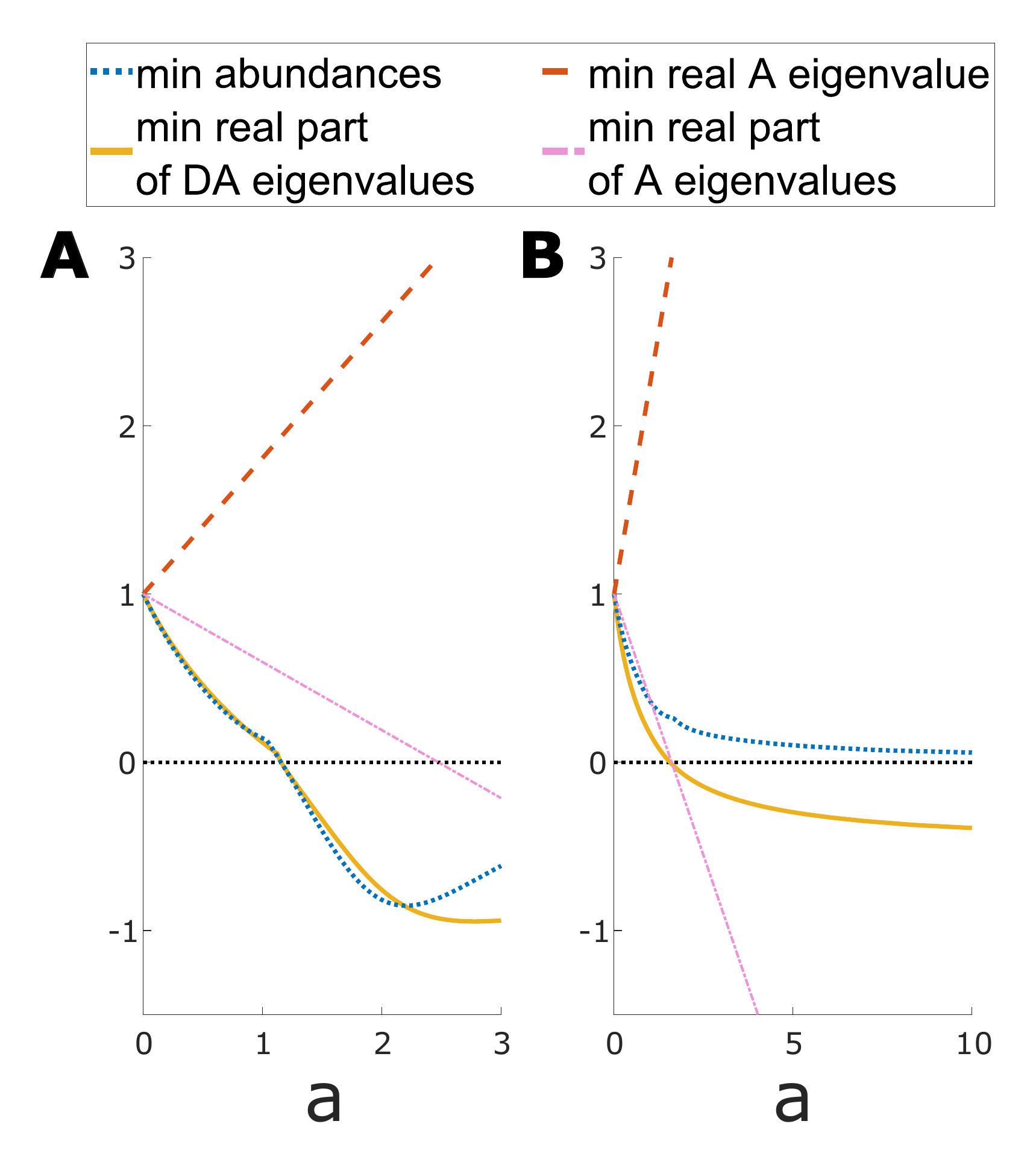}
\par\end{centering}
\caption{\textbf{\label{fig:appendix asymmetric matrices}Asymmetric interactions
with no divergences in the abundances. }Matrices are of the form $\mathbf{A}=\mathbf{I}+a\mathbf{M}$,
with values shown as a function of $a$. The minimal abundance is
shown in a dotted blue line. The minimum of the real parts of $\mathbf{DA}$
eigenvalues is in a full yellow line. The dashed red line shows the
smallest of the real valued eigenvalues of $\mathbf{A}$ and is linear
with $a$. The minimum of the real parts of $\mathbf{A}$ eigenvalues
is shown in a thin dashed-dotted pink line. In both cases, the smallest
real eigenvalue of $\mathbf{M}$ is positive and so $\mathbf{A}$
is always invertible. As a result, the abundances do not diverge \textbf{(A)}
feasibility and Lyapunov stability break together\textbf{ (B)} Only
Lyapunov stability breaks. The abundances are positive for any positive
finite $a$.}
\end{figure}

In Fig. \ref{fig:appendix asymmetric matrices} we show two examples
of cases where there are no divergences in the abundances. In \ref{fig:appendix asymmetric matrices}A,
feasibility is lost for some positive $a$, and in \ref{fig:appendix asymmetric matrices}B
it is never lost. The matrices used are
\begin{align*}
\mathbf{M}_{\text{A}} & =\left(\begin{array}{ccc}
0 & 0.2 & 0.3\\
0.9 & 0 & 0\\
0.8 & 0.7 & 0
\end{array}\right)\\
\mathbf{M}_{\text{B}} & =\left(\begin{array}{ccc}
0 & 0.9 & 0.2\\
0.3 & 0 & 1\\
0.7 & 0.6 & 0
\end{array}\right)
\end{align*}

The $a$-dependent abundances for $\mathbf{M}_{\text{B}}$ are given
by $N_{i}^{*}\left(a\right)=\left(\mathbf{I}+a\mathbf{M}_{\text{B}}\right)^{-1}\vec{K}$
(for all $K_{i}=1$). No species ever has an abundance of zero, since
$N_{i}\left(a\right)=0$ has no real solution for $a$; however $N_{i}^{*}$
\textit{tends} to zero as $a\rightarrow\infty$. The behaviors are
not fine-tuned, and remain unchanged when adding random perturbations
to $M_{ij}$, $i\neq j$, from a normal distribution $\mathcal{N}\left(0,10^{-4}\right)$.

\section{\label{subsec:Lyapunov-stability-real leading EV}Lyapunov stability
breaking before feasibility for asymmetric matrices with a real leading
eigenvalue}

Here consider competitive asymmetric matrices $\mathbf{A}$ with a
leading real eigenvalue, and check whether Lyapunov stability breaks
down before feasibility. Taking linear trajectories in $\left\{ A_{ij}\right\} $
space that cross the origin, this seems to be a rare occurrence, although
there may be other types of trajectories where this would be more
common. We consider matrices of sizes $3\times3$ and $4\times4$
and of the form $\mathbf{A}=\mathbf{I}+a\mathbf{M}$, with $\mathbf{I}$
the identity matrix and $\mathbf{M}$ having zeros along the diagonal
and off-diagonal elements taken from a uniform distribution over $\left[0,1\right]$.
We randomly generate 10000 matrices $\mathbf{M}$ of each size, then
see whether Lyapunov stability breaks before feasibility as $a$ is
increased. For $3\times3$ matrices, we find that for matrices $\mathbf{M}$
that have a real leading eigenvalue, this occurs with probability
of about $0.36\%$, in contrast to probability of about $6.8\%$ for
matrices where the leading eigenvalue is a complex conjugate pair.
For $4\times4$ matrices, for matrices $\mathbf{M}$ that have a real
leading eigenvalue, this occurs with probability of about $0.43\%$,
in contrast to probability of about $12.7\%$ for matrices where the
leading eigenvalue is a complex conjugate pair.

\bibliographystyle{unsrt}
\bibliography{Feasibility_instability}

\end{document}